\documentclass[conference]{IEEEtran}
\IEEEoverridecommandlockouts
\usepackage{cite}
\usepackage{amsmath,amssymb,amsfonts}
\usepackage{algorithmic}
\usepackage{graphicx}
\usepackage{textcomp}
\usepackage{xcolor}
\usepackage{graphicx}    
\usepackage{lipsum}      
\usepackage[nolist]{acronym}
\usepackage{comment}
\usepackage{soul}
\usepackage{multirow}
\usepackage{array}

\usepackage[absolute,overlay]{textpos}

\usepackage[caption=false,font=small,captionskip=2pt]{subfig}
\usepackage{tikz}
\usetikzlibrary{positioning, calc, arrows.meta, fit,decorations.pathreplacing}
\usepackage{pgfplots}
\pgfplotsset{compat=1.18}

\usepackage{xcolor}
\usepackage[normalem]{ulem}
\newcommand{\addedAmorosa}[1]{\textcolor{black}{#1}}
\newcommand{\removedAmorosa}[1]{}
\newcommand{\changedAmorosa}[2]{\textcolor{black}{#2}}

\newcommand{\addedCippo}[1]{\textcolor{black}{#1}}

\def\BibTeX{{\rm B\kern-.05em{\sc i\kern-.025em b}\kern-.08em
    T\kern-.1667em\lower.7ex\hbox{E}\kern-.125emX}}

\IEEEoverridecommandlockouts
    
\begin{document}


\title{ML-Based Channel Quality Prediction\\for URLLC Services of Connected Vehicles}
\title{Machine Learning-Based Channel Quality Prediction for URLLC in Connected and Automated Vehicles}
\title{Predictive Channel Quality Estimation for URLLC Services in Connected Vehicles Using Deep Learning}
\title{Deep Learning-Based Channel Quality Prediction for Adaptive URLLC Vehicular Services}
\title{Proactive Channel Quality Prediction for 5G/6G URLLC Vehicular Communications}
\title{Channel Quality Prediction for URLLC Vehicular Services Using DNN and LSTM Models}
\title{Enabling Proactive URLLC Adaptation in Connected Vehicles Through ML-Based Channel Prediction}
\title{Learning to Predict Wireless Channel Quality for Reliable Low-Latency Vehicular Services}
\title{Towards Predictive URLLC for Connected Vehicles: A Deep Learning Approach}
\title{Deep Neural Prediction of Channel Quality for Adaptive 5G/6G Vehicular URLLC}
\title{AI-Driven Channel Prediction for Reliable and Low-Latency Connected Vehicle Services}
\title{Proactive URLLC Adaptation for Connected Vehicles Through ML-Based Channel Prediction}

\author{\IEEEauthorblockN{
Andrea Giovannini\IEEEauthorrefmark{1},
Lorenzo Mario Amorosa\IEEEauthorrefmark{2}\IEEEauthorrefmark{1}, 
Vittorio Todisco\IEEEauthorrefmark{2}\IEEEauthorrefmark{1},
Claudia Campolo\IEEEauthorrefmark{3}\IEEEauthorrefmark{1},\\
Antonella Molinaro\IEEEauthorrefmark{3}\IEEEauthorrefmark{1},
Su Hongjia\IEEEauthorrefmark{4}, Alessandro Bazzi\IEEEauthorrefmark{2}\IEEEauthorrefmark{1}
}
\IEEEauthorblockA{\IEEEauthorrefmark{1}National Laboratory of Wireless Communications (WiLab), CNIT, 40136 Bologna, Italy}
\IEEEauthorblockA{\IEEEauthorrefmark{3}DEI, Universit\`a di Bologna, 40136 Bologna, Italy}
\IEEEauthorblockA{\IEEEauthorrefmark{2}Universit\`a Mediterranea di Reggio Calabria, 89124 Reggio Calabria, Italy}\IEEEauthorblockA{\IEEEauthorrefmark{4}Heisenberg Research Center, Huawei Technologies Duesseldorf GmbH, 80992 Munich, Germany}
\thanks{©2026 IEEE.  Personal use of this material is permitted.  Permission from IEEE must be obtained for all other uses, in any current or future media, including reprinting/republishing this material for advertising or promotional purposes, creating new collective works, for resale or redistribution to servers or lists, or reuse of any copyrighted component of this work in other works.}
}

\IEEEpubid{\makebox[\columnwidth]{979-8-3195-0489-0/26/\$31.00~\copyright~2026 European Union\hfill}\hspace{\columnsep}\makebox[\columnwidth]{}}

\maketitle
\begin{textblock*}{3cm}(17cm,1cm)
\fontsize{10}{12}\selectfont (Special Session)
\end{textblock*}

\bstctlcite{IEEEexample:BSTcontrol}

\begin{acronym} 
\acro{3GPP}{Third Generation Partnership Project}
\acro{5G}{fifth generation}
\acro{5GC}{5G core}
\acro{5QI}{5G Quality of Service Indicator}
\acro{6G}{sixth generation}
\acro{5GAA}{5G Automotive Association}
\acro{AF}{Application Function}
\acro{AR}{allocation round}
\acro{ARFCN}{absolute radio-frequency channel number}
\acro{AS}{Application Server}
\acro{API}{Application Programming Interface}
\acro{BS}{base station}
\acro{C-V2X}{cellular-vehicle-to-everything}
\acro{CAD}{connected and automated driving}
\acro{CAV}{connected and automated vehicle}
\acro{CDF}{cumulative distribution function}
\acro{CN}{core network}
\acro{CSI}{channel state information}
\acro{DDQ-N}{double deep Q-learning network}
\acro{DL}{downlink}
\acro{DNN}{deep neural network}
\acro{EED}{end-to-end delay}
\acro{eMBB}{enhanced mobile broadband}
\acro{FoV}{field of view}
\acro{GNSS}{global navigation satellite system}
\acro{HD}{high-definition}
\acro{IQN}{in-advance QoS notification}
\acro{KPI}{key performance indicator}  
\acro{LDPC}{low-density parity check}
\acro{LGBM}{light gradient boosted machine}
\acro{LoA}{Level of Automation}
\acro{LR}{linear regression}
\acro{LSTM}{long short-term memory}
\acro{LTE}{long term evolution}  
\acro{MAC}{medium access control}
\acro{MAE}{mean absolute error}
\acro{MEC}{multi-access edge computing}
\acro{MCS}{modulation and coding scheme}
\acro{MIMO}{multiple input multiple output}
\acro{ML}{machine learning}
\acro{MLP}{multilayer perceptron} 
\acro{MNO}{mobile network operator}
\acro{MSE}{mean square error}
\acro{MU-MIMO}{multi-user multiple-input multiple-output}
\acro{NWDAF}{network data analytics function}
\acro{NEF}{network exposure function}
\acro{NF}{network function}
\acro{NN}{neural network}
\acro{NR}{new radio}
\acro{OAM}{operations, administration and maintenance}
\acro{OFDM}{orthogonal frequency-division multiplexing}
\acro{OSM}{OpenStreetMap}
\acro{PER}{packet error rate}
\acro{PF}{prediction function}
\acro{PHY}{physical}
\acro{pQoS}{predictive quality of service}
\acro{PRR}{packet reception ratio}
\acro{QoE}{quality of experience}
\acro{QoS}{quality of service}
\acro{RAN}{radio access network}
\acro{REM}{radio environmental map}
\acro{RF}{random forest}
\acro{RL}{reinforcement learning}
\acro{RMSE}{root mean square error}
\acro{RNN}{recurrent neural network}
\acro{RRM}{radio resource management}
\acro{RSRP}{reference signal received power}
\acro{RSRQ}{reference signal received quality}
\acro{RSSI}{received signal strength indicator}
\acro{RSU}{road side unit} 
\acro{SGD}{stochastic gradient descent}
\acro{SINR}{signal-to-interference-plus-noise ratio}
\acro{SLA}{service level agreement}
\acro{SLR}{service level requirement}
\acro{SNR}{signal-to-noise ratio}
\acro{SoTA}{state-of-the-art}
\acro{SUMO}{Simulation of Urban MObility}
\acro{SVM}{support vector machine}
\acro{TCP}{transmission control protocol}
\acro{ToD}{tele-operated driving}
\acro{TTI}{transmission time interval}
\acro{UPF}{User Plane Function}
\acro{URLLC}{ultra-reliable and ultra-low latency communications}
\acro{USRP}{universal software radio peripheral}
\acro{V2N}{vehicle-to-network}
\acro{V2I}{vehicle-to-infrastructure} 
\acro{V2P}{vehicle-to-pedestrian}
\acro{V2V}{vehicle-to-vehicle} 
\acro{V2X}{vehicle-to-everything} 
\acro{VRU}{vulnerable road user}
\acro{VUE}{vehicular user equipment}
\acro{WP}{work package}
\end{acronym}

\removedAmorosa{Fare attenzione alla policy di submission: https://coinsconf.com/authors-guide-2/  * double blind review: alla prima submission vanno omessi gli autori. Target: 6 pagine. Revisionare le figure.}







\begin{abstract}
Connected and automated vehicles (CAVs) are expected to increasingly rely on 5G and future 6G ultra-reliable and low-latency communication (URLLC) services to support safety-critical and time-sensitive applications. Since wireless link conditions can vary rapidly in urban vehicular environments, proactively adapting service parameters based on future channel conditions is essential to maintain service continuity and reliability.
In this paper, we investigate the use of machine learning (ML) techniques for channel quality prediction in vehicular URLLC scenarios. Specifically, we evaluate deep neural network (DNN) and long short-term memory (LSTM) models to forecast future channel conditions and enable proactive service adaptation with minimized performance degradation. The analysis is conducted using realistic simulations combining the SUMO traffic simulator and the Sionna-RT ray-tracing framework in a real urban environment reconstructed from OpenStreetMap data.
Results show that ML-based prediction significantly outperforms approaches relying solely on past channel measurements and achieves performance close to the ideal case in which future channel conditions are perfectly known in advance. These findings demonstrate the potential of ML-driven prediction techniques to enhance the reliability and robustness of URLLC services for connected vehicular systems.
\end{abstract}

\section{\addedAmorosa{Introduction}}

\label{use}


For conventional vehicles driven by humans, the best path to reach a destination is either the fastest one or the shortest one. When considering a \ac{CAV}, whose safety and comfort significantly rely on the quality of the network connectivity assisting the driving to enhance its perception of the surrounding environment, other options may be preferred. 
Indeed, as detailed by the \ac{5GAA} in \cite{5GAA_usecases2}, it is envisioned that future \acp{CAV} will exploit a service called \textit{infrastructure-assisted environment perception} to improve\removedAmorosa{its} awareness of the surrounding\addedAmorosa{s} and control\removedAmorosa{its} movements; in such \addedAmorosa{a} service, the vehicle continuously receives \removedAmorosa{from the network}environmental updates collected by other vehicles or by \removedAmorosa{cameras and other}sensors installed along the roads. In addition to \removedAmorosa{the list of} static and dynamic objects, \removedAmorosa{
and especially those that are beyond the range of the vehicles' own sensors,}
\removedAmorosa{also}trajectories and actuation commands \removedAmorosa{may} need to be transmitted. This service \removedAmorosa{will require}\addedAmorosa{requires} a large throughput in \ac{DL}, from a few hundreds~kbps to more than 80~Mbps \cite{5GAA_usecases2}, \cite{5GAA_qos}\removedAmorosa{and, overall, a high level of \ac{QoS}}. If the required \removedAmorosa{downlink}throughput is not available, then the amount of \addedAmorosa{delivered}information \removedAmorosa{the network can transmit will} decrease\addedAmorosa{s}, \removedAmorosa{thereby limiting the horizon or level of detail of the known environment and potentially raising safety concerns.} \addedAmorosa{violating the strict \ac{SLA} required for safe autonomous navigation.} 
\removedAmorosa{In order to reproduce and investigate this scenario, we use the concept of \textit{\ac{CAD} Modes}, defined already in our previous works \cite{10978586} and \cite{11230595}. The \ac{CAD} Mode indicates the driving conditions that the level of awareness allow.} 
\addedAmorosa{To model the interplay between network performance and vehicular automation, we leverage the concept of \ac{CAD} Modes \cite{11230595}.} CAD Mode~0 \removedAmorosa{is used to }indicate\addedAmorosa{s} that the \ac{CAV} \removedAmorosa{has no support from the network and }relies solely on its onboard sensors\removedAmorosa{, therefore suggesting driving cautiously}. Higher CAD Modes are enabled when {high} throughput conditions are experienced, \removedAmorosa{which guarantee a better awareness and, consequently, a more confident driving, allowing for example shorter following distances and higher average speeds. }\addedAmorosa{supporting advanced automated maneuvers.} 

Because wireless link conditions can vary rapidly in dynamic urban vehicular environments, proactively adapting service parameters based on future channel conditions is essential to maintain service continuity and reliability \cite{10978586}. To navigate these rapid fluctuations, the network must rely on continuously accurate short-term prediction of the channel to determine the most suited \ac{CAD} Mode and to satisfy the associated \ac{SLA}. In this framework, \ac{ML}-driven techniques become helpful in enhancing the communication with respect to non-\ac{ML} procedures.
\addedAmorosa{To address this challenge, this work evaluates data-driven architectures, specifically \ac{DNN} and \ac{LSTM} models, in forecasting short-term link quality. We demonstrate that these ML models provide significantly more robust short-term channel estimation compared to reactive baselines, effectively safeguarding the stringent latency and reliability budgets demanded by \ac{URLLC} applications.}
The main contributions of this paper are summarized as follows:
\begin{itemize}
    \item We propose a short-term channel quality estimation framework utilizing \ac{DNN} and \ac{LSTM} architectures to enable proactive \ac{URLLC} service adaptation for connected and automated vehicles.
    \item We evaluate our framework using a highly realistic ray-tracing-backed vehicular simulation combining SUMO \cite{sumo} and Sionna-RT \cite{sionna} modeled on a real-world urban area in Bologna.
    \item Through extensive benchmarks, we demonstrate that the proposed \ac{ML} approaches achieve robust reliability, reducing global \ac{SLA} failure rates close to the perfect-channel-knowledge bound.
\end{itemize}

\addedAmorosa{\section{Related Work}\label{subsec:relatedwork}
Recent literature has increasingly explored data-driven techniques for channel prediction to support URLLC. For instance, studies such as \cite{reyhanoglu2023machine} and \cite{9860971} demonstrated the efficacy of DNNs in predicting \ac{SINR} and link delays in interference-limited scenarios.
Expanding into dynamic vehicular mobility, deep learning approaches have been proposed to predict wireless channel parameters for edge computing in intelligent connected vehicles, demonstrating significant accuracy improvements over conventional auto-regressive models \cite{Liu2019_DeepLearningV2X}.} 

\addedAmorosa{Time-series forecasting models, particularly LSTMs, have also been utilized to capture temporal dependencies and anticipate sudden throughput drops \cite{9604941, skocaj2023vehicle}. 
In \ac{V2I} environments, LSTMs have successfully been employed as network filter processes to predict channel conditions for optimal radio access technology (RAT) selection, significantly reducing queue backlogs \cite{AlObaidi2022_LSTM_RAT}. Furthermore, recent advancements have benchmarked \acp{RNN} directly against classical Wiener filters for 2D-channel estimation in highly time- and frequency-selective \ac{V2X} conditions, proving that data-driven equalizers exhibit superior resilience against Doppler spread and system parameter mismatches \cite{Fischer2021_WienerRNN}.} 

\addedAmorosa{Looking toward future 6G deployments, the state-of-the-art is shifting toward integrating spatial-frequency \ac{ML}-based channel estimation \cite{Ye2025_GNN_MIMO} and multimodal collaborative perception frameworks \cite{Gharsallah2025_Multimodal6G}. However, most existing works focus predominantly on uplink channels, static environments, or single-metric predictions. This paper extends the state-of-the-art by applying predictive DNN and LSTM architectures to highly dynamic urban vehicular \ac{DL} communications, evaluating their impact directly on SLA failure rates and proactive \ac{QoS} adaptation.}

\section{System Model}\label{sec:model}

We assume a multi-cell scenario where the target \ac{CAV} moves towards a certain destination while supported by an infrastructure-assisted environment perception service. 
Depending on the available resources, the level of service changes, following the concept of \ac{CAD} Modes with corresponding \ac{SLA} \cite{11230595}. The highest CAD Modes allow the CAV to receive more information and therefore to drive with a longer perception horizon and with higher safety levels. 
Since the service is of type \ac{URLLC}, only small delays and virtually no losses are tolerable.

Given this target service, \textit{the network needs to allocate the radio resources to the CAV with the highest possible \ac{CAD} Mode that is compatible with the corresponding \ac{SLA} and the radio link conditions}. \addedCippo{Thus, a reliable channel estimation is of paramount importance to ensure that the service requirements are met.}
Assuming a dedicated network slice is available \cite{campolo20175g} and a low service penetration, this study focuses on a single CAV without inter-cell interference. These simplifying assumptions will be relaxed in future works.




\newlength{\lowerboxwidth}
\setlength{\lowerboxwidth}{3.15cm}

\newlength{\headerboxwidth}
\setlength{\headerboxwidth}{9.62cm}

\begin{figure}[t]
\centering
\hspace*{1.3cm}%
\resizebox{0.98\columnwidth}{!}{%
\begin{tikzpicture}[
    font=\sffamily\scriptsize,
    >=Latex,
    flow/.style={
        rounded corners=4pt,
        draw=#1!80!black,
        fill=#1!38,
        thick,
        align=center,
        minimum width=\lowerboxwidth,
        minimum height=0.82cm,
        text width=3.85cm
    },
    smallflow/.style={
        rounded corners=3pt,
        draw=gray!70!black,
        fill=gray!35,
        thick,
        align=center,
        minimum width=\lowerboxwidth,
        minimum height=0.82cm,
        text width=3.85cm
    },
    blueflow/.style={
        rounded corners=3pt,
        draw=blue!75!black,
        fill=blue!18,
        thick,
        align=center,
        minimum width=\lowerboxwidth,
        minimum height=0.82cm,
        text width=3.85cm
    },
    ttiflow/.style={
        rounded corners=4pt,
        draw=purple!80!black,
        fill=purple!38,
        thick,
        align=center,
        minimum width=2.70cm,
        minimum height=0.74cm,
        text width=2.40cm
    },
    outcome/.style={
        rounded corners=3pt,
        draw=gray!70!black,
        fill=gray!35,
        thick,
        align=center,
        minimum width=2.95cm,
        minimum height=0.68cm,
        text width=3.85cm
    },
    bracelabel/.style={
        align=center,
        font=\sffamily\fontsize{5.6pt}{6.5pt}\selectfont,
        text width=1.55cm
    },
    arrowlabel/.style={
        align=center,
        font=\sffamily\fontsize{5.6pt}{6.5pt}\selectfont,
        fill=white,
        inner sep=1pt
    },
    cyclelabel/.style={
        align=center,
        font=\sffamily\fontsize{6pt}{6.8pt}\selectfont,
        fill=white,
        inner sep=1pt
    }
]

\node[
    minimum width=\headerboxwidth,
    minimum height=0pt,
    inner sep=0pt,
    outer sep=0pt
] (scaleref) {};

\node[
    blueflow
] (sensing) at ($(scaleref.center)+(-1.40cm,1.95cm)$)
    {Short-term channel prediction};

\node[
    smallflow,
    below=0.38cm of sensing
] (cadmode)
    {CAD mode selection\\[1pt]
    {\fontsize{5.6pt}{6.5pt}\selectfont estimated spectral eff.}};

\node[
    flow=green,
    below=0.38cm of cadmode
] (allocation)
    {Resource allocation\\[1pt]
    {\fontsize{5.6pt}{6.5pt}\selectfont new packets + retries}};

\node[
    ttiflow,
    below=0.55cm of allocation,
    xshift=0.45cm
] (position)
    {Update\\position};

\node[
    ttiflow,
    below=0.28cm of position
] (transmission)
    {Transmission over\\actual channel};

\node[
    ttiflow,
    below=0.28cm of transmission
] (failure)
    {Failure\\evaluation};

\node[
    outcome,
    below=0.60cm of failure,
    xshift=-0.45cm
] (outcome)
    {Retransmission buffer\\or SLA failure\\[1pt]
    {\fontsize{5.6pt}{6.5pt}\selectfont according to delay budget}};

\draw[->, thick]
    (sensing.south) -- (cadmode.north);

\draw[->, thick]
    (cadmode.south) -- (allocation.north);

\draw[
    ->,
    thick,
    rounded corners=5pt
]
    (allocation.south) -- ++(0,-0.28cm) -| (position.north);

\draw[->, thick]
    (position.south) -- (transmission.north);

\draw[->, thick]
    (transmission.south) -- (failure.north);

\draw[
    ->,
    thick,
    rounded corners=5pt
]
    (failure.south) -- ++(0,-0.30cm) -| (outcome.north);


\draw[
    decorate,
    decoration={brace, amplitude=5pt, mirror},
    thick
]
    ($(sensing.north west)+(-0.70cm,0.10cm)$) --
    ($(outcome.south west)+(-0.70cm,-0.10cm)$)
    node[
        bracelabel,
        midway,
        left=7pt
    ]
    {Repeated at\\every allocation\\round\\[5pt]{\large$\circlearrowright$}};

\draw[
    decorate,
    decoration={brace, amplitude=5pt, mirror},
    thick
]
    ($(position.north west)+(-0.28cm,0.08cm)$) --
    ($(failure.south west)+(-0.28cm,-0.08cm)$)
    node[
        bracelabel,
        midway,
        left=0pt
    ]
    {Repeated\\every TTI\\[5pt]{\large$\circlearrowright$}};

\end{tikzpicture}%
}
\caption{Allocation-rounds and time transmission intervals.}
\label{fig:simulator_online}
\end{figure}
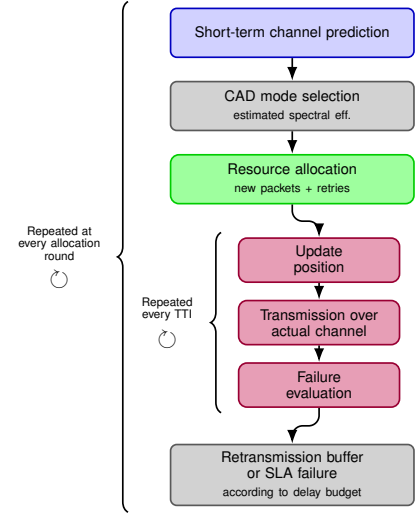


\subsection{Radio Access and Allocation}

While moving, the CAV is connected to a given cell. The network allocates the resources in that cell and the service performs data transmission, followed by link-level retransmission in case of losses. We assume the infrastructure-assisted environment perception service to be hosted at the network edge
, which implies 
that the delay and losses introduced outside the radio link are negligible.

Focusing on the radio link, the allocation and transmission process is organized in \textit{\acp{AR}} and \textit{\acp{TTI}}, as represented in Fig.~\ref{fig:simulator_online}. The \ac{AR} represents the granularity for radio resource allocation and includes a given number of TTIs, which in turn represent the granularity of the transmission. At the beginning of the \ac{AR}, the channel quality is estimated 
(thus defining the \ac{MCS}), the CAD Mode is selected, and the resources are allocated. Then, during each TTI the signal is transmitted in the allocated radio resources and the correctness of the reception is verified. 

For the sake of resource allocation, an \textit{estimation of the channel} is needed, which we denote as $\widehat{L}$. 
How to accurately estimate $\widehat{L}$ is the subject of this work. 
Denoting the actual power loss by $L_\text{actual}$, the objective of short-term channel estimation is to map a historical sequence of past power loss and contextual measurements $\mathbf{x}$ spanning a past window of $N$ TTIs to an estimated channel loss value $\hat{L}(t)$:
\begin{equation}
    \hat{L}(t) = f\big(\{L_{actual}(t-1, \tau), \mathbf{x}(t-1, \tau)\}_{\tau=1}^{N}\big)
\end{equation}
where $f(\cdot)$ represents the estimation framework (as detailed in Sec.~\ref{ml-pred}). 

Given 
$\widehat{L}$, the network estimates the maximum throughput that it can support $\widehat{T}_\text{DL, max}$, which is computed through the following equations: 
\begin{equation}
\widehat{\gamma}=\frac{P_\text{T} \, G_\text{T} \, G_\text{R}}{\widehat{L} \, P_\text{N}}=\frac{P_\text{T} \, G_\text{T} \, G_\text{R}}{\widehat{L} \, F_\text{R} \, k \, T_\text{0} \, B_\text{W}},  \label{eq:gamma}
\end{equation}
\begin{equation}
\widehat{\eta}=\alpha \, \log_2(1+\widehat{\gamma}), \label{eq:eta}
\end{equation}
\begin{equation}
\widehat{T}_\text{DL, max}=\widehat{\eta}_\text{max} \,  B_\text{W}, \label{eq:throughput}
\end{equation}
where $\widehat{\gamma}$ is the estimated \ac{SNR} when assuming that all the bandwidth is used, $P_\text{T}$ is the power transmitted before the BS antenna, $G_\text{T}$ and $G_\text{R}$ are the transmitting and receiving antenna gains, respectively, $P_\text{N}$ is the noise power, $F_\text{R}$ is the receiver noise figure, $k$ is the Boltzmann constant, $T_\text{0}$ is the reference temperature of 290 K, $\widehat{\eta}$ is the estimated channel efficiency, and  
$\alpha$ is a parameter that accounts for the protocol efficiency. 
To be noted that the channel efficiency corresponds in the real system to a certain \ac{MCS}, with a lower \ac{MCS} providing higher reliability at the cost of lower efficiency and viceversa. Therefore, $\widehat{\eta}$ represents the maximum \ac{MCS} that reliably guarantees  $\widehat{T}_\text{DL, max}$ if the SNR equals $\widehat{\gamma}$.
The maximum estimated throughput $\widehat{T}_\text{DL, max}$ is then used to select the highest CAD Mode that can be supported. 
The latter is referred to as \textit{actual CAD Mode} and its corresponding throughput is denoted as target throughput $T^*_\text{DL}$ (with $T^*_\text{DL}\leq\widehat{T}_\text{DL, max}$). 

Even if a single CAV and no inter-cell interference are considered, we 
schedule the minimum amount of resources required by the CAV to use the given CAD Mode. 
In particular, the network assumes the \ac{MCS} corresponding to the channel efficiency $\widehat{\eta}$ and allocates the bandwidth $B_\text{actual}\leq B_\text{W}$ computed as: 
\begin{equation}
    B_\text{actual} = T^*_\text{DL}/\widehat{\eta}\;.
\end{equation}


\subsection{Data Transmission and Retransmission}\label{subsec:correctness}


The maximum throughput $\overline{T}_\text{DL}$ that can be supported without losses, assuming the allocated bandwidth $B_\text{actual}$ and the actual channel conditions $\gamma_\text{actual}$, can be derived as: 
\begin{equation}
\gamma_\text{actual}=\frac{P_\text{T} \, G_\text{T} \, G_\text{R}}{L_\text{actual} \, F_\text{R} \, k \, T_\text{0} \, B_\text{actual}}, \label{eq:gammaActual}
\end{equation}
\begin{equation}
\overline{T}_\text{DL}=\alpha B_\text{actual} \log_2(1+\gamma_\text{actual})\;.
\end{equation}



If $T^*_\text{DL}<\overline{T}_\text{DL}$, we assume no losses in the transmissions, otherwise part of the data needs retransmission. 
Specifically, we assume that the number of lost bits $b_\text{err}$ needing retransmission equals:
\begin{equation}
    b_\text{err} = 
\begin{cases}
    0 & \text{if } R_\text{DL} \leq \overline{T}_\text{DL} \\
    ( T^*_\text{DL} - \overline{T}_\text{DL} ) \cdot T_\text{TTI} & \text{otherwise, }
\end{cases}
\end{equation}
where $T_\text{TTI}$ is the duration of the 
\ac{TTI}. 
The bits $b_\text{err}$ 
are buffered and retransmitted during the next \ac{AR} with higher priority than the new data. The retransmitted bits that are not correctly received within the allowed maximum delay (or \textit{delay budget}) are discarded, 
triggering an \ac{SLA} failure event.

\section{Case Study Scenario and Main Simulation Settings}\label{subsec:scenario}

\begin{figure}[t]
    \centering
    \includegraphics[
        width=\columnwidth,
        trim=0.25cm 0.2cm 0.05cm 0.5cm,
        clip
    ]{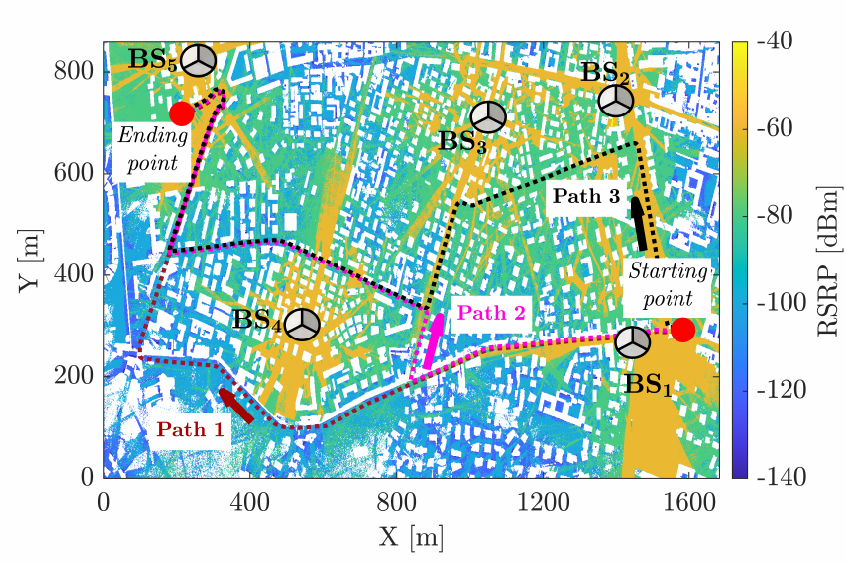}
    \caption{Urban evaluation scenario in Bologna, Italy, featuring the layout of the 15 base station sectors, the three distinct vehicular routes, and the ray-traced RSRP coverage map.}
    \label{fig:scenario}
\end{figure}

The considered scenario is a portion of Bologna close to the Engineering Faculty. The road map was derived from the OpenStreetMap project \cite{osm}. The area, shown in Fig.~\ref{fig:scenario}, includes three distinct vehicular routes (namely Path~1, Path~2, and Path~3), each connecting the same origin and destination. Mobility patterns were generated in SUMO \cite{sumo}, considering road speed limits (up to 50~km/h) and free-flow traffic conditions. The three routes are comparable in length (between 2100 and 2400~m) and 
travel time (between 170 and 190~s). 
Five tri-sector \acp{BS} (i.e., 15 cells) are located according to the real deployment as indicated in the LTEItaly database \cite{lteitaly}. 
Each BS has antennas at a height of 25~m, with three cells oriented at 30°, 150°, and 270°, using single-element antennas with 3GPP TR 38.901 radiation patterns \cite{38.901}, as natively supported by Sionna RT. Vehicles are equipped with 1.5~m-high dipole antennas. Environmental features, such as building geometry and road layout, were imported from OpenStreetMap to enable accurate modeling. 
%
Concerning the online simulation, the adopted \ac{TTI} is 0.5~ms and the \ac{AR} is 5~ms. The maximum tolerated delay is set to 20~ms \cite{5gaa-20ms}. Regarding the handover mechanism, a combination of a dwell-time-based mechanism and an \ac{SNR} threshold is implemented, where the handover is initiated if a neighboring cell's SNR exceeds that of the serving cell by more than 3~dB and this condition persists for at least 1~s.
From a service perspective, we assume three CAD Modes beyond zero, as detailed in Table~\ref{tab:exCADmodes}. 

\begin{table}[t]\footnotesize
\centering
\caption{Operational mapping and requirements of \ac{CAD} modes.}
\label{tab:exCADmodes}
{\renewcommand{\arraystretch}{1.35}
\begin{tabular}{|>{\centering\arraybackslash}m{0.55cm}|
                >{\centering\arraybackslash}m{1.88cm}|
                >{\raggedright\arraybackslash}m{5cm}|}
\hline
\textbf{CAD Mode} &
\textbf{DL throughput req. [Mbps]} &
\multicolumn{1}{>{\centering\arraybackslash}m{5cm}|}{\textbf{Supported perception and operation}} \\
\hline \hline
0 &
No requirement &
No network-assisted perception; the \ac{CAV} relies only on onboard sensors. \\
\hline
1 &
$1 < T_\text{DL} \leq 20$ &
Basic network-assisted perception, suitable only for conservative automated operation. \\
\hline
2 &
$20 < T_\text{DL} \leq 50$ &
Intermediate network-assisted perception, supporting automated operation with reduced confidence or tighter constraints. \\
\hline
3 &
$T_\text{DL} > 50$ &
Extended and accurate network-assisted perception, enabling the highest supported automation level. \\
\hline
\end{tabular}
}
\end{table}

\section{Short-Term Channel Prediction}
\label{ml-pred}

\addedAmorosa{\subsection{ML-Assisted Short-Term Channel Prediction Models}\label{subsec:MLmodels}}
\addedAmorosa{To achieve accurate short-term channel estimation, we benchmarked two distinct architectures: a feed-forward \ac{DNN} and a recurrent \ac{LSTM} network.}
%
\addedAmorosa{Both architectures are trained to forecast imminent channel conditions based on the 
dataset generated via the Sionna RT and SUMO integration.}  

\textit{Dataset:} Both models are trained on a set of engineered features derived from \ac{RSRP} measurements and spectral efficiency experienced by a single vehicle traveling in our simulation framework over each of the 3 available paths of the case study scenario recalled in Section~\ref{subsec:scenario}. Each sample of the dataset is composed of a series of  consecutive \ac{RSRP} values over the past $N$ \acp{TTI} 
that are first min-max normalized. Then, the derivative, the minimum and the maximum values, the mean and the standard deviation in this interval are computed. 
The same is done for a series of $N$ values of the spectral efficiency, which serves as contextual information. The size of the dataset is in the order of $10^5$ samples.
The training set is made from the samples related to Path 1 and Path 2 (i.e., about 66\% of the total), while the test set is made from the samples related to Path 3 (i.e., about 33\% of the total). These inputs provide both instantaneous and statistical context over a specific time window, allowing the network to potentially analyze trends and signal variability for better predictions.

\textit{DNN settings:} The architecture chosen for the \ac{DNN} is a fully connected \ac{MLP} with three hidden layers, each containing 256 neurons. The learning rate  is set to \(1\times 10^{-5}\), the mini-batch size is 64, 
\addedAmorosa{and the model is trained for a single epoch}. 





\textit{LSTM settings:} Regarding the \ac{LSTM} model, it 
\addedAmorosa{comprises 2 recurrent layers} with 100 neurons \addedAmorosa{each}. The learning rate  is set to \(1\times 10^{-5}\), the mini-batch size is 64, the number of epochs is 1. 

For both networks, the low number of epochs is justified by the \addedAmorosa{dense} size of the dataset, which \addedAmorosa{facilitates rapid convergence}.

\textit{Loss functions:} To optimize the networks for URLLC constraints, we evaluate three distinct objective functions: \ac{MAE}, \ac{MSE}, and a custom asymmetric loss function (denoted as MAE+MSE), which 
applies standard \ac{MAE} penalties for channel underestimations, but switches to an \ac{MSE} penalty for channel overestimations. This asymmetric behavior strictly punishes optimistic channel predictions, which are catastrophic in URLLC settings because they lead to severe resource starvation, packet drops, and fatal retransmission loops.


\textit{Output:} Both models aim to provide a specific, single-value prediction: the worst case \ac{RSRP} in the next \ac{AR}. Considering \acp{AR} of a length of 10 \acp{TTI}, this corresponds to predicting the worst case over the next 10 \acp{TTI}. 
This aims at determining the worst channel conditions in the \ac{AR} in order to avoid and limit retransmissions. 
Starting from this prediction, the spectral efficiency and the achievable bit rate can be derived. We chose to predict the worst case in order to limit retransmissions and data loss and to have a system that is more forgiving with imprecise predictions. In fact, the dataset is subject to fast channel variations and predicting the best case or an average case would have led to a network that is more susceptible to noise and prone to make riskier predictions.  


\smallskip
\subsection{Benchmarks}
The following schemes are considered as benchmarks for the proposed \ac{ML}-based solutions.
\begin{itemize}
\item Ideal channel knowledge in each TTI, where the scheduler knows in advance the exact channel in each TTI of the AR; this is used as an unrealistic upper-bound to the performance, indicated as ``Ideal per TTI”;
\item Ideal channel knowledge of the worst case in the AR, where the scheduler knows in advance the channel of the worst TTI in the AR; this benchmark also sets the upper-bound which the scheduler should target with its estimation, and is indicated as ``Ideal per AR”;
\item Estimation of the channel in the next AR equal to the mean of the last $N$ \acp{TTI}; this scheme gives a benchmark of standard approaches based on the average past channel and is indicated as ``Past average”;
\item Estimation of the channel in the next AR equal to the worst case of the last $N$ \acp{TTI}; this scheme is a benchmark of non-ML solutions, where a conservative approach is used to cope with the URLLC requirements; this scheme is denoted as ``Past minimum”.
\end{itemize}

These approaches set benchmarks with two ideal solutions (the first two items) and two history-based approaches. Given the URLLC requirements, a conservative approach needs to be used and therefore the main benchmarks are the ``Ideal per AR” and ``Past minimum”, while the other two are provided as references of what would be achieved neglecting the requirements.

\subsection{Metrics of Interest}\label{sec:outputmetrics}

Results are derived in terms of the following metrics to quantify the effectiveness of the proposed short-term channel estimation:
\begin{itemize}
\item  \textbf{\textit{CAD score}}: it measures the effectiveness in terms of the selected CAD Mode for 
the chosen path, calculated as:
\begin{equation}
S =\sum_{m=0}^{M}w_m CAD_{m}^{\%}\;,
\end{equation}
where $CAD_{m}^{\%}$ is the percentage of time spent in each CAD Mode, $M$ is the number of CAD Modes, and $w_m$ defines the 
weights that are set as $w_0=0$, $w_1=0.02$, $w_2=0.4$, and $w_3=1$ to capture the highest relevance of the higher CAD Modes. 
This setting is derived from the ratio between the throughput associated to the CAD Mode and the maximum achievable throughput. This ensures that the CAD score $S$ varies between $0$ and $1$. In addition, this set choice allows to interpret $S$ as the average throughput experienced by the \ac{CAV} on the path, expressed as a percentage of the maximum achievable throughput. For instance, a path score of $0\%$ and $100\%$ corresponds to the limit cases where the \ac{CAV} spends all the time either in CAD Mode 0  or in CAD Mode 3 (i.e., $0\%$ and $100\%$ of the maximum throughput on average).
\item  \textbf{\textit{\ac{SLA} failure rate}}: it measures the portion of time when the agreed \ac{SLA} is not satisfied, calculated as:
\begin{equation}\label{eq:Rsf}
R_{sf}=\frac{T_{sf}}{T_{path}},
\end{equation}
where $T_{sf}$ is the time during which the \ac{SLA} is not satisfied, and $T_{path}$ is the overall travel time; 
the \ac{SLA} failure rate is given as an overall average or separately per each CAD Mode; in the latter case, both the numerator and denominator of 
\eqref{eq:Rsf} are restricted to the given CAD Mode, and this allows to assess the impact of failures that are associated to different safety risks. 
\end{itemize}

\section{Results}\label{sec:results}



\changedAmorosa{This section describes the results obtained applying the proposed \ac{DNN} and \ac{LSTM} architectures for channel estimation. Due to space constraints, we report the entirety of the obtained results only for the \ac{MAE} loss function case (see Fig.~\ref{fig:DNN-metrics-MAE} and Fig.~\ref{fig:DNN-fail-cad-MAE}). The results related to the other loss functions are summarized in Table~\ref{tab:res_ML_summary}, with data associated to different numbers of past \ac{TTI} samples, namely 10, 50 and 100 samples.}{This section evaluates the performance of the proposed \ac{DNN} and \ac{LSTM} short-term predictors. 
Comprehensive performance metrics across all loss functions for structural benchmarks ($N \in \{10, 50, 100\}$) are presented in Table~\ref{tab:res_ML_summary_perc}.} 

\subsection{Channel Estimation using DNN}

\removedAmorosa{The first group of results refer to the DNN specified in Section~\ref{ml-pred}, with different loss functions.}

\subsubsection{DNN with MAE} 

\removedAmorosa{Results for the \ac{DNN} (alongside the \ac{LSTM}) when considering  \ac{MAE} as loss function are reported in Fig.~\ref{fig:DNN-metrics-MAE}, and compare it against the aforementioned benchmarks when varying the number of past \acp{TTI}, $N$.} 

Figs.~\ref{fig:DNN-path-MAE} and~\ref{fig:DNN-tot-fail-MAE} report the CAD score \addedAmorosa{($S$)}, the \addedAmorosa{total} SLA failure rate \addedAmorosa{($R_{sf}$)}, and {\ac{CAD}-aggregated metrics against the past window size $N$. While \textit{Ideal per TTI} represents a theoretical performance bound under continuous feedback, \textit{Ideal per AR} serves as the realistic optimization target for the scheduler.} When considering history-based \changedAmorosa{channel prediction}{baselines}, we can observe that 
\textit{Past average} is too optimistic. \changedAmorosa{When using it, resources are allocated in such a way that a larger score than the \textit{Ideal per AR} is achieved, but this comes at the cost of frequent failures. Such failures get larger as $N$ increases.}{ It registers an inflated path score outperforming the \textit{Ideal per AR} limit, but triggers severe SLA violations that scale monotonically with $N$.} 
%
The \ac{DNN} \changedAmorosa{allows to achieve}{stably tracks} a score close to the \textit{Ideal per AR} \changedAmorosa{and such trend is not sensitive to the chosen $N$ (see Fig.~\ref{fig:DNN-path-MAE}). Interestingly, for values of $N$ greater than 50 past \acp{TTI}, the SLA failure rate is below 0.005 (see Fig.~\ref{fig:DNN-tot-fail-MAE}).}{independently of $N$ (Fig.~\ref{fig:DNN-path-MAE}), achieving an aggregated SLA failure rate below $0.5\%$ for $N \geq 50$ (Fig.~\ref{fig:DNN-tot-fail-MAE}).} 

\changedAmorosa{The SLA failure rates per each individual \ac{CAD} Mode are then shown in Fig.~\ref{fig:DNN-fail-cad-MAE}. As observable, the behavior is consistent with the observations made for Fig.~\ref{fig:DNN-metrics-MAE}.}{Specific \ac{CAD} Mode configurations are detailed in Fig.~\ref{fig:DNN-fail-cad-MAE}. The baseline behaviors mirror the aggregated case}\changedAmorosa{ The curves related to \textit{Past minimum} decrease with a higher number of past \acp{TTI}, as the prediction tends to be more pessimistic and cautious. On the other hand, those related to \textit{Past average} increase with a higher number of past \acp{TTI} as the prediction tends to be more optimistic and riskier.}{, where expanded context penalizes the performance of \textit{Past minimum} and escalates the risk profile of \textit{Past average}.} A notable exception is observed for \ac{CAD} Mode~3, where the \textit{Past average} is stable and below 0.2\%\changedAmorosa{. This means that when the channel can support the highest \ac{CAD} Mode, usually it is also either more predictable or can better tolerate imprecise estimations (i.e., even if the channel has strong variations, the worst channel conditions still allow the highest \ac{CAD} Mode).}{, indicating that high-SNR conditions are inherently more resilient to variance and estimation mismatch.} The \ac{DNN} \changedAmorosa{allows to achieve lower SLA failure rates than the benchmarks for each \ac{CAD} Mode with a past window of at least 50 \acp{TTI}, with the worst SLA failure rate for \ac{CAD} Mode~2 and the best one for \ac{CAD} Mode~3.}{consistently outperforms the empirical baselines across all operating modes when $N \geq 50$, hitting its optimal reliability bounds in \ac{CAD} Mode~3.}

\removedAmorosa{The considerations done for the \ac{SLA} failure rates of the single \ac{CAD} Modes are equivalently observed also in the other loss functions cases. }

\subsubsection{DNN with MSE} 
\changedAmorosa{Corresponding results when the DNN is used considering  \ac{MSE} as loss function are 
summarized in Table~\ref{tab:res_ML_summary_perc}, which reports the metrics of interest values for a past window of 10, 50 and 100 \acp{TTI}, respectively. 
%
The observations that can be made are comparable to those stated in the \ac{MAE} case. {We notice that increasing the past window size leads to a decrease in the path score, but also in a decrease of the \ac{SLA} failure rates. The improvement is greater from 10 \ac{TTI} samples to 50 \ac{TTI} samples than from 50 \ac{TTI} samples to 100 \ac{TTI} samples.}}{Evaluating the \ac{MSE} configuration in Table~\ref{tab:res_ML_summary_perc} reveals identical trends regarding context scaling; structural gains in reliability materialize sharply when moving from $10$ to $50$ \acp{TTI}, entering a saturation plateau at $100$ \acp{TTI}.} 
Comparing the \addedAmorosa{the DNN-MSE framework to} \removedAmorosa{values with those related to the}\ac{MAE} \addedAmorosa{estimates}, we observe a slightly lower score \addedAmorosa{($56.85\%$ vs. $58.14\%$ at $50$ \acp{TTI})}\removedAmorosa{and a slightly higher used bandwidth}. This behavior is due to the fact that the loss function now is harsher with respect to outliers and the network tends towards a more cautious behavior.

\subsubsection{DNN with a Combination of MAE and MSE} 

\changedAmorosa{In Table~\ref{tab:res_ML_summary}, results for the \ac{DNN} when considering  an asymmetric loss function mixing \ac{MAE} and \ac{MSE} are reported. In particular, the loss function corresponds to the \ac{MAE} for an underestimation and to the \ac{MSE} for an overestimation. The aim of this function is to penalize the overestimation, since overestimating the output can easily lead to retransmissions, while an underestimation does not.}{The mixed asymmetric configuration (Table~\ref{tab:res_ML_summary_perc}) 
explicitly punishes channel overestimations, which are critical catalysts for packet drops and URLLC retransmissions.}  
\addedAmorosa{This structural penalty induces a highly cautious behavior: at $50$ \acp{TTI}, the path score decreases significantly ($52.24\%$), but yields the lowest overall \ac{DNN} SLA failure rate 
($0.24\%$), validating the intended design objective.}


\definecolor{mycolor1}{rgb}{0.00000,0.44700,0.74100}
\definecolor{mycolor2}{rgb}{0.85098,0.32549,0.09804}
\definecolor{mycolor3}{rgb}{0.92941,0.69412,0.12549}
\definecolor{mycolor4}{rgb}{0.49412,0.18431,0.55686}
\definecolor{mycolor5}{rgb}{0.46667,0.67451,0.18824}
\definecolor{mycolor6}{rgb}{0.30196,0.74510,0.93333}

\begin{figure}[!t]
    \centering

    \begin{minipage}{\columnwidth}
        \centering


\definecolor{mycolor1}{rgb}{0.00000,0.44700,0.74100}
\definecolor{mycolor2}{rgb}{0.85098,0.32549,0.09804}
\definecolor{mycolor3}{rgb}{0.92941,0.69412,0.12549}
\definecolor{mycolor4}{rgb}{0.49412,0.18431,0.55686}
\definecolor{mycolor5}{rgb}{0.46667,0.67451,0.18824}
\definecolor{mycolor6}{rgb}{0.30196,0.74510,0.93333}

\begin{tikzpicture}
\begin{axis}[
    width=0.52\linewidth,
    height=0.40\linewidth,
    scale only axis,
    xmin=0,
    xmax=200,
    ymin=40,
    ymax=80,
    xtick={0,50,100,150,200},
    ytick={40,50,60,70,80},
    xlabel={Number of past TTIs},
    ylabel={CAD Score $(\%)$},
    font=\small,
    label style={font=\footnotesize},
    tick label style={font=\footnotesize},
    grid=both,
    major grid style={line width=0.25pt, draw=gray!35},
    minor grid style={line width=0.15pt, draw=gray!18},
    tick align=inside,
    tick style={gray},
    axis line style={black},
    axis lines=box,
    legend style={
    draw=none,
    fill=none,
    font=\footnotesize,
    at={(0,0)},
    anchor=south west,
    opacity=0,
    text opacity=0
},
    every axis plot/.append style={
        line width=1.2pt,
        mark size=1.2pt
    }
]

\addplot[color=mycolor1, mark=*]
table[row sep=crcr]{
10 60.3222164948454\\
20 58.3923711340206\\
30 58.1656701030928\\
40 58.3877835051547\\
50 58.1395360824742\\
60 58.6161340206186\\
70 58.1113917525773\\
80 57.8995360824742\\
90 58.1444845360825\\
100 58.0944329896907\\
110 57.974381443299\\
120 58.2839175257732\\
130 58.7659793814433\\
140 58.3038144329897\\
150 58.8029381443299\\
160 58.4325773195876\\
170 57.6518556701031\\
180 57.8571649484536\\
190 57.7513402061856\\
200 57.5474742268042\\
};
\addlegendentry{DNN}

\addplot[color=mycolor2, mark=square*]
table[row sep=crcr]{
10 59.0092783505155\\
20 56.5949484536083\\
30 55.4832989690722\\
40 55.1017010309278\\
50 54.2731443298969\\
60 54.2568041237114\\
70 54.4572680412371\\
80 56.5369587628866\\
90 56.0780412371134\\
100 54.4883505154639\\
110 55.8696391752578\\
120 55.075\\
130 54.928762886598\\
140 54.3730412371134\\
150 56.3040721649485\\
160 56.6737113402062\\
170 55.2746391752578\\
180 56.240206185567\\
190 57.7465979381444\\
200 57.5611855670103\\
};
\addlegendentry{LSTM}

\addplot[color=mycolor3, mark=triangle*]
table[row sep=crcr]{
10 60.1122680412371\\
20 56.0465463917526\\
30 53.6213402061856\\
40 51.8956185567011\\
50 50.5517525773196\\
60 49.4729896907217\\
70 48.5429896907217\\
80 47.7449484536083\\
90 47.0188659793815\\
100 46.3544845360825\\
110 45.7475257731959\\
120 45.1700515463918\\
130 44.6275773195877\\
140 44.1201546391753\\
150 43.6421134020619\\
160 43.1909793814433\\
170 42.7599484536083\\
180	42.3562371134021\\
190 41.9592268041237\\
200 41.5772164948454\\
};
\addlegendentry{Past minimum}

\addplot[color=mycolor4, mark=diamond*]
table[row sep=crcr]{
10	73.0301030927835\\
20	73.1297422680412\\
30	73.1427319587629\\
40	73.1271649484536\\
50	73.1081443298969\\
60	73.0714948453608\\
70	73.0412886597938\\
80	73.0172164948454\\
90	73.0030927835052\\
100	72.9774742268041\\
110	72.974587628866\\
120	72.9296907216495\\
130	72.929175257732\\
140	72.8946907216495\\
150	72.9024742268041\\
160	72.8637113402062\\
170	72.8628350515464\\
180	72.8436082474227\\
190	72.7907216494845\\
200	72.7735567010309\\
};
\addlegendentry{Past average}

\addplot[color=mycolor5, dashed, mark=none]
table[row sep=crcr]{
10 60.4273711340206\\
200 60.4273711340206\\
};
\addlegendentry{Ideal per AR}

\addplot[color=mycolor6, densely dashed, mark=none]
table[row sep=crcr]{
10 75.6914432989691\\
200 75.6914432989691\\
};
\addlegendentry{Ideal per TTI}

\end{axis}
\end{tikzpicture}%
        \hspace{-0.01\columnwidth}%
        \raisebox{4em}{%
            \begin{tikzpicture}
                \begin{axis}[
                    hide axis,
                    scale only axis,
                    width=5pt,
                    height=5pt,
                    xmin=0, xmax=1,
                    ymin=0, ymax=1,
                    legend columns=1,
                    legend style={
                        draw=black,
                        fill=white,
                        font=\footnotesize,
                        cells={anchor=west},
                        legend cell align=left,
                        inner sep=1.2pt,
                        row sep=0pt,
                        at={(0,0.5)},
                        anchor=west
                    }
                ]

                \addlegendimage{color=mycolor1, line width=1.35pt, mark=*}
                \addlegendentry{DNN}

                \addlegendimage{color=mycolor2, line width=1.35pt, mark=square*}
                \addlegendentry{LSTM}

                \addlegendimage{color=mycolor3, line width=1.35pt, mark=triangle*}
                \addlegendentry{Past min.}

                \addlegendimage{color=mycolor4, line width=1.35pt, mark=diamond*}
                \addlegendentry{Past avg.}

                \addlegendimage{color=mycolor5, line width=1.35pt, dashed}
                \addlegendentry{Ideal per AR}

                \addlegendimage{color=mycolor6, line width=1.35pt, densely dashed}
                \addlegendentry{Ideal per TTI}

                \end{axis}
            \end{tikzpicture}
        }%

    \end{minipage}

    \caption{\small Score comparison of DNN, LSTM, and baseline methods using MAE.
    }
    \label{fig:DNN-path-MAE}
\end{figure}
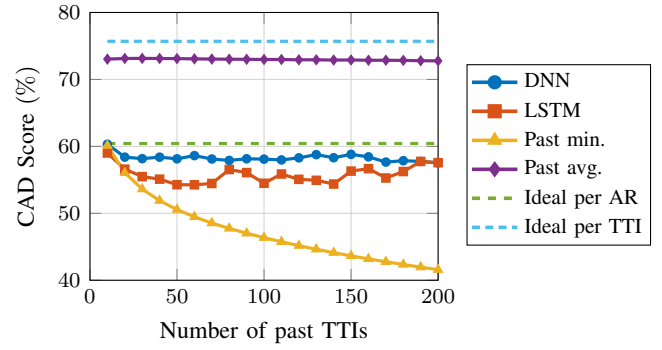

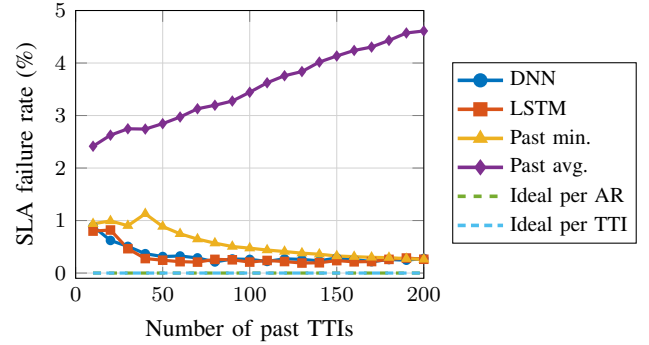
\begin{figure}[!t]
    \centering

    \begin{minipage}{\columnwidth}
        \centering


\definecolor{mycolor1}{rgb}{0.00000,0.44700,0.74100}
\definecolor{mycolor2}{rgb}{0.85098,0.32549,0.09804}
\definecolor{mycolor3}{rgb}{0.92941,0.69412,0.12549}
\definecolor{mycolor4}{rgb}{0.49412,0.18431,0.55686}
\definecolor{mycolor5}{rgb}{0.46667,0.67451,0.18824}
\definecolor{mycolor6}{rgb}{0.30196,0.74510,0.93333}

\begin{tikzpicture}
\begin{axis}[
    width=0.52\linewidth,
    height=0.40\linewidth,
    scale only axis,
    xmin=0,
    xmax=200,
    ymin=-0.1,
    ymax=5,
    xtick={0,50,100,150,200},
    ytick={0,1,2,3,4,5},
    xlabel={Number of past TTIs},
    ylabel={SLA failure rate (\%)},
    y filter/.expression={y*100},
    font=\small,
    label style={font=\footnotesize},
    tick label style={font=\footnotesize},
    grid=both,
    major grid style={line width=0.25pt, draw=gray!35},
    minor grid style={line width=0.15pt, draw=gray!18},
    tick align=inside,
    tick style={gray},
    axis line style={black},
    axis lines=box,
    legend style={
    draw=none,
    fill=none,
    font=\footnotesize,
    at={(0,0)},
    anchor=south west,
    opacity=0,
    text opacity=0
},
    every axis plot/.append style={
        line width=1.2pt,
        mark size=1.2pt
    }
]

\addplot[color=mycolor1, mark=*]
table[row sep=crcr]{
10 0.00889175257731267\\
20 0.00621134020619252\\
30 0.00507731958762747\\
40 0.00368556701030798\\
50 0.00314432989691227\\
60 0.00324742268040268\\
70 0.00288659793815782\\
80 0.00211340206186605\\
90 0.00268041237112016\\
100 0.00260309278351656\\
110 0.00221649484535646\\
120 0.00270618556700697\\
130 0.00260309278351656\\
140 0.0023195876288753\\
150 0.00288659793815782\\
160 0.00260309278351656\\
170 0.00216494845361126\\
180 0.00257731958762975\\
190 0.00244845360825252\\
200 0.00278350515463899\\
};
\addlegendentry{DNN}

\addplot[color=mycolor2, mark=square*]
table[row sep=crcr]{
10 0.00798969072164368\\
20 0.00819587628865293\\
30 0.00463917525772217\\
40 0.00275773195875217\\
50 0.00244845360825252\\
60 0.00216494845361126\\
70 0.00208762886597924\\
80 0.00260309278351656\\
90 0.00252577319588454\\
100 0.00208762886597924\\
110 0.00239690721650732\\
120 0.00219072164949807\\
130 0.00190721649485681\\
140 0.00195876288660202\\
150 0.0023711340206205\\
160 0.00213917525772445\\
170 0.00219072164949807\\
180 0.00260309278351656\\
190 0.00286082474227101\\
200 0.00262886597937495\\
};
\addlegendentry{LSTM}

\addplot[color=mycolor3, mark=triangle*]
table[row sep=crcr]{
10 0.00938144329896318\\
20 0.00992268041235889\\
30 0.0090206185566899\\
40 0.01128865979382\\
50 0.00886597938145428\\
60 0.00747422680413479\\
70 0.00646907216494697\\
80 0.0056958762886552\\
90 0.00505154639174066\\
100 0.004742268041241\\
110 0.00438144329896772\\
120 0.00409793814432646\\
130 0.00378865979382681\\
140 0.00358247422681757\\
150 0.00324742268040268\\
160	0.00311855670103093\\
170 0.00298969072164823\\
180 0.00293814432990303\\
190 0.00273195876289378\\
200 0.00260309278351656\\
};
\addlegendentry{Past minimum}

\addplot[color=mycolor4, mark=diamond*]
table[row sep=crcr]{
10 0.0241494845360819\\
20 0.026262886597948\\
30 0.0274742268041166\\
40 0.0274226804123714\\
50 0.0284536082474176\\
60 0.0297164948453599\\
70 0.0312886597938018\\
80 0.0319587628866032\\
90 0.0327577319587533\\
100 0.0344329896907141\\
110 0.036237113402052\\
120 0.0375773195876263\\
130 0.0383762886598049\\
140 0.0401804123711429\\
150 0.0413402061855663\\
160 0.0423969072164994\\
170 0.0430412371134139\\
180 0.0443041237113277\\
190 0.0457216494845341\\
200 0.0461082474226941\\
};
\addlegendentry{Past average}

\addplot[color=mycolor5, dashed, mark=none]
table[row sep=crcr]{
10 0\\
200 0\\
};
\addlegendentry{Ideal per AR}

\addplot[color=mycolor6, densely dashed, mark=none]
table[row sep=crcr]{
10 0\\
200 0\\
};
\addlegendentry{Ideal per TTI}

\end{axis}
\end{tikzpicture}%
        \hspace{-0.01\columnwidth}%
        \raisebox{4em}{%
            \begin{tikzpicture}
                \begin{axis}[
                    hide axis,
                    scale only axis,
                    width=5pt,
                    height=5pt,
                    xmin=0, xmax=1,
                    ymin=0, ymax=1,
                    legend columns=1,
                    legend style={
                        draw=black,
                        fill=white,
                        font=\footnotesize,
                        cells={anchor=west},
                        legend cell align=left,
                        inner sep=1.2pt,
                        row sep=0pt,
                        at={(0,0.5)},
                        anchor=west
                    }
                ]

                \addlegendimage{color=mycolor1, line width=1.35pt, mark=*}
                \addlegendentry{DNN}

                \addlegendimage{color=mycolor2, line width=1.35pt, mark=square*}
                \addlegendentry{LSTM}

                \addlegendimage{color=mycolor3, line width=1.35pt, mark=triangle*}
                \addlegendentry{Past min.}

                \addlegendimage{color=mycolor4, line width=1.35pt, mark=diamond*}
                \addlegendentry{Past avg.}

                \addlegendimage{color=mycolor5, line width=1.35pt, dashed}
                \addlegendentry{Ideal per AR}

                \addlegendimage{color=mycolor6, line width=1.35pt, densely dashed}
                \addlegendentry{Ideal per TTI}

                \end{axis}
            \end{tikzpicture}
        }%

    \end{minipage}

    \caption{\small SLA failure rate comparison of DNN, LSTM, and baseline methods using MAE.
    }
    \label{fig:DNN-tot-fail-MAE}
\end{figure}

\begin{figure*}[!t]
    \centering

    \begin{tikzpicture}
        \begin{axis}[
            hide axis,
            xmin=0, xmax=1,
            ymin=0, ymax=1,
            legend columns=6,
            legend style={
                draw=black,
                fill=white,
                line width=0.3pt,
                font=\footnotesize,
                at={(0.5,0.5)},
                anchor=center,
                legend cell align=left,
                /tikz/every even column/.append style={column sep=0.6em}
            }
        ]

        \addlegendimage{color=mycolor1, line width=1.35pt, mark=*}
        \addlegendentry{DNN}

        \addlegendimage{color=mycolor2, line width=1.35pt, mark=square*}
        \addlegendentry{LSTM}

        \addlegendimage{color=mycolor3, line width=1.35pt, mark=triangle*}
        \addlegendentry{Past minimum}

        \addlegendimage{color=mycolor4, line width=1.35pt, mark=diamond*}
        \addlegendentry{Past average}

        \addlegendimage{color=mycolor5, line width=1.35pt, dashed}
        \addlegendentry{Ideal per AR}

        \addlegendimage{color=mycolor6, line width=1.35pt, densely dashed}
        \addlegendentry{Ideal per TTI}

        \end{axis}
    \end{tikzpicture}

    \vspace{-0.6em}

    \makebox[\textwidth][c]{%
        \subfloat[CAD Mode 1.]{
            \label{fig:fail_rate_CAD_1_MAE}
            \begin{minipage}[t]{0.315\textwidth}
                \centering
\begin{tikzpicture}

\begin{axis}[
    width=\linewidth,
    height=0.84\linewidth,
    xmin=0,
    xmax=200,
    ymin=-0.5,
    ymax=30,
    xtick={0,50,100,150,200},
    ytick={0,10,20,30},
    xlabel={Number of past TTIs},
    ylabel={Failure rate (\%)},
    y filter/.expression={y*100},
    font=\footnotesize,
    label style={font=\footnotesize},
    tick label style={font=\footnotesize},
    grid=both,
    major grid style={line width=0.25pt, draw=gray!35},
    minor grid style={line width=0.15pt, draw=gray!18},
    axis lines=box,
    tick align=inside,
    tick style={draw=none},
    axis line style={black},
    every axis plot/.append style={
        line width=1pt,
        mark size=1.2pt
    }
]
\addplot [color=mycolor1, mark=*]
  table[row sep=crcr]{
10	0.0805152979066008\\
20	0.0461460446247486\\
30	0.0427782888684476\\
40	0.0345655304848833\\
50	0.0296519123334633\\
60	0.0327082424770992\\
70	0.0295234078447777\\
80	0.0194253338729311\\
90	0.0268262484523234\\
100	0.0274771024146503\\
110	0.0244416350610948\\
120	0.0308896210873115\\
130	0.0295546558704416\\
140	0.0261162594776749\\
160	0.0278001611603546\\
150	0.0315962248666324\\
170	0.0239425379090221\\
180	0.0277219766974781\\
190	0.0264195583596347\\
200	0.0310225967062365\\
};

\addplot [color=mycolor2, mark=square*]
  table[row sep=crcr]{
10	0.0529999999999973\\
20	0.0233990147783345\\
30	0.0161290322580783\\
40	0.0153536370500831\\
50	0.0152889349572547\\
60	0.0145607107601222\\
70	0.0147936053447779\\
80	0.020014825796892\\
90	0.0183440753594368\\
100	0.0144113336590124\\
110	0.017212921480791\\
120	0.0158438576349056\\
130	0.0135331077815408\\
140	0.0129262313349727\\
150	0.0178527757397831\\
160	0.0161538461538555\\
170	0.0151233131689139\\
180	0.0185354691075474\\
190	0.0225409836065467\\
200	0.0206047631790227\\
};

\addplot [color=mycolor3, mark=triangle*]
  table[row sep=crcr]{
10	0.0700636942675033\\
20	0.0868878357030098\\
30	0.0753012048192829\\
40	0.0949640287769853\\
50	0.0787292817679486\\
60	0.0630026809651554\\
70	0.0620155038759549\\
80	0.0542929292929273\\
90	0.0483870967742064\\
100	0.0471584038694175\\
110	0.0415676959619873\\
120	0.038416763678697\\
130	0.0365714285714286\\
140	0.034792368125693\\
150	0.0308710033075954\\
160	0.0291891891891892\\
170	0.0275715800636362\\
180	0.0260145681581605\\
190	0.0234933605720187\\
200	0.0230460921843587\\
};

\addplot [color=mycolor4, mark=diamond*]
  table[row sep=crcr]{
10	0.192329708070986\\
20	0.201448151021452\\
30	0.213485941776554\\
40	0.207378828068471\\
50	0.215124471582897\\
60	0.222324936445574\\
70	0.229452825940115\\
80	0.228019538188278\\
90	0.232964601769908\\
100	0.243072223434439\\
110	0.253382005583006\\
120	0.260093497662552\\
130	0.261725293132315\\
140	0.269651330720023\\
150	0.274087740877405\\
160	0.279013344116464\\
170	0.281783490660786\\
180	0.286139793486882\\
190	0.291976516634037\\
200	0.291804555187866\\
};

\addplot [color=mycolor5, dashed, mark=none]
  table[row sep=crcr]{
10	0\\
200	0\\
};

\addplot [color=mycolor6, densely dashed, mark=none]
  table[row sep=crcr]{
10	0\\
200	0\\
};

\end{axis}
\end{tikzpicture}%
                \vspace{-0.4em}
            \end{minipage}
        }
        \hspace{-0.025\textwidth}
        \subfloat[CAD Mode 2.]{
            \label{fig:fail_rate_CAD_2_MAE}
            \begin{minipage}[t]{0.315\textwidth}
                \centering
\begin{tikzpicture}

\begin{axis}[
    width=\linewidth,
    height=0.84\linewidth,
    xmin=0,
    xmax=200,
    ymin=-0.15,
    ymax=10,
    xtick={0,50,100,150,200},
    ytick={0,2.5,7.5,10},
    xlabel={Number of past TTIs},
    ylabel={Failure rate (\%)},
    y filter/.expression={y*100},
    font=\footnotesize,
    label style={font=\footnotesize},
    tick label style={font=\footnotesize},
    grid=both,
    major grid style={line width=0.25pt, draw=gray!35},
    minor grid style={line width=0.15pt, draw=gray!18},
    axis lines=box,
    tick align=inside,
    tick style={draw=none},
    axis line style={black},
    every axis plot/.append style={
        line width=1pt,
        mark size=1.2pt
    }
]
\addplot [color=mycolor1, mark=*]
  table[row sep=crcr]{
10	0.0797342192691133\\
20	0.0414250207125235\\
30	0.0332906530089758\\
40	0.026164645820046\\
50	0.018366054464849\\
60	0.0179233621755372\\
70	0.0139064475347652\\
80	0.00982197667281071\\
90	0.0117719950433752\\
100	0.00838825644098051\\
110	0.00790754257906201\\
120	0.00906892382104729\\
130	0.00791717417783389\\
140	0.00664251207729194\\
150	0.00927643784785914\\
160	0.0085731781996401\\
170	0.00575705238918545\\
180	0.00937316930287579\\
190	0.00775193798449436\\
200	0.006049606775548\\
};

\addplot [color=mycolor2, mark=square*]
  table[row sep=crcr]{
10	0.0648252536640257\\
20	0.0625434329395489\\
30	0.0585058505850498\\
40	0.0145038167939049\\
50	0.011049723756912\\
60	0.00710479573712064\\
70	0.0059978189749188\\
80	0.0041649312786376\\
90	0.00548754748839997\\
100	0.00606343283581623\\
110	0.00406669377795765\\
120	0.00256410256409367\\
130	0.00375156315132585\\
140	0.00429184549355455\\
150	0.00300042863267436\\
160	0.0033396946564892\\
170	0.003464703334771\\
180	0.00331674958539452\\
190	0.00419463087249028\\
200	0.00518358531317631\\
};

\addplot [color=mycolor3, mark=triangle*]
  table[row sep=crcr]{
10	0.056677524429972\\
20	0.0570228091236515\\
30	0.0584562996594684\\
40	0.0648854961831944\\
50	0.0495049504950487\\
60	0.0405202601300516\\
70	0.0349006301502754\\
80	0.0277908619877394\\
90	0.0223438212494216\\
100	0.0217294900221816\\
110	0.0203639514731435\\
120	0.0186282811176852\\
130	0.0182194616977256\\
140	0.0162140251317453\\
150	0.0139165009940427\\
160	0.0132605304212063\\
170	0.0134099616858236\\
180	0.0136105860113389\\
190	0.0123180291153415\\
200	0.011049723756912\\
};

\addplot [color=mycolor4, mark=diamond*]
  table[row sep=crcr]{
10	0.0753911806543499\\
20	0.0703571428571479\\
30	0.0585365853658573\\
40	0.0587207270185388\\
50	0.0531315974665745\\
60	0.0547419804742049\\
70	0.0558446063128599\\
80	0.062197650310992\\
90	0.0627356873500275\\
100	0.0637355146557601\\
110	0.06475548060709\\
120	0.0666891209161236\\
130	0.0690235690235568\\
140	0.0716893461666075\\
150	0.0778403444849403\\
160	0.076209410205422\\
170	0.0749260598093997\\
180	0.0784570120954697\\
190	0.0797007156799054\\
200	0.0812641083521441\\
};

\addplot [color=mycolor5, dashed, mark=none]
  table[row sep=crcr]{
10	0\\
200	0\\
};

\addplot [color=mycolor6, densely dashed, mark=none]
  table[row sep=crcr]{
10	0\\
200	0\\
};

\end{axis}
\end{tikzpicture}%
                \vspace{-0.4em}
            \end{minipage}
        }
        \hspace{-0.025\textwidth}
        \subfloat[CAD Mode 3.]{
            \label{fig:fail_rate_CAD_3_MAE}
            \begin{minipage}[t]{0.315\textwidth}
                \centering
\begin{tikzpicture}

\begin{axis}[
    width=\linewidth,
    height=0.84\linewidth,
    xmin=0,
    xmax=200,
    ymin=-0.03,
    ymax=2,
    xtick={0,50,100,150,200},
    ytick={0,0.5,1,1.5,2},
    xlabel={Number of past TTIs},
    ylabel={Failure rate (\%)},
    y filter/.expression={y*100},
    font=\footnotesize,
    label style={font=\footnotesize},
    tick label style={font=\footnotesize},
    grid=both,
    major grid style={line width=0.25pt, draw=gray!35},
    minor grid style={line width=0.15pt, draw=gray!18},
    axis lines=box,
    tick align=inside,
    tick style={draw=none},
    axis line style={black},
    every axis plot/.append style={
        line width=1pt,
        mark size=1.2pt
    }
]
\addplot [color=mycolor1, mark=*]
  table[row sep=crcr]{
10	0.00868578412115539\\
20	0.00451793620675289\\
30	0.00237442922374953\\
40	0.0013645046848012\\
50	0.00109689213894626\\
60	0.000997732426299081\\
70	0.000914620203960305\\
80	0.000827053850400716\\
90	0.000914662032386104\\
100	0.000962199312709799\\
110	0.000688420762770647\\
120	0.000684806428049671\\
130	0.000678886625934183\\
140	0.000775830595102889\\
150	0.000904159132005589\\
160	0.000819336337571031\\
170	0.000647428782826864\\
180	0.00069073494196914\\
190	0.000691690491549934\\
200	0.00078649086282212\\
};

\addplot [color=mycolor2, mark=square*]
  table[row sep=crcr]{
10	0.00640620770548139\\
20	0.00978143866709047\\
30	0.00366039170944532\\
40	0.00129957643434864\\
50	0.00107453355474263\\
60	0.000640551860072947\\
70	0.000393855848756175\\
80	0.000478583393146437\\
90	0.000530631934395842\\
100	0.000445500445493963\\
110	0.000485224901751735\\
120	0.000491497100171046\\
130	0.000444005920078934\\
140	0.000398505603982358\\
150	0.000576064519236752\\
160	0.000616903146209324\\
170	0.000587170328316233\\
180	0.000577784197588471\\
190	0.000608215589039673\\
200	0.000609384521624179\\
};

\addplot [color=mycolor3, mark=triangle*]
  table[row sep=crcr]{
10	0.0102656738775977\\
20	0.0111548867897682\\
30	0.00980733807935508\\
40	0.0130493088508388\\
50	0.0101954120645757\\
60	0.00881344866982658\\
70	0.00728020451262523\\
80	0.00673839184597114\\
90	0.00622478386168268\\
100	0.00562489013887557\\
110	0.00523497917905047\\
120	0.00495049504951339\\
130	0.00434729365662179\\
140	0.00421993297754852\\
150	0.00396002262868933\\
160	0.00381825124094348\\
170	0.00354198866563138\\
180	0.00345119489483636\\
190	0.00329185594839032\\
200	0.00319446293090664\\
};

\addplot [color=mycolor4, mark=diamond*]
  table[row sep=crcr]{
10	0.00195276518920195\\
20	0.00158222025979171\\
30	0.00147324223786427\\
40	0.00132616223385185\\
50	0.00136310050103816\\
60	0.00125377977727226\\
70	0.00125470514430503\\
80	0.00121847653508667\\
90	0.00129303975174366\\
100	0.00129390018483377\\
110	0.00125762899943993\\
120	0.00133259300389454\\
130	0.00125865324102392\\
140	0.00133427226566596\\
150	0.00118601979170307\\
160	0.00129797886148708\\
170	0.0014469095495997\\
180	0.00141059430565351\\
190	0.00137490245623439\\
200	0.00141300710222936\\
};

\addplot [color=mycolor5, dashed, mark=none]
  table[row sep=crcr]{
10	0\\
200	0\\
};

\addplot [color=mycolor6, densely dashed, mark=none]
  table[row sep=crcr]{
10	0\\
200	0\\
};

\end{axis}
\end{tikzpicture}%
                \vspace{-0.4em}
            \end{minipage}
        }
    }
    
    \caption{\small SLA failure rate comparison for the three CAD modes using MAE. }
    \label{fig:DNN-fail-cad-MAE}
\end{figure*}
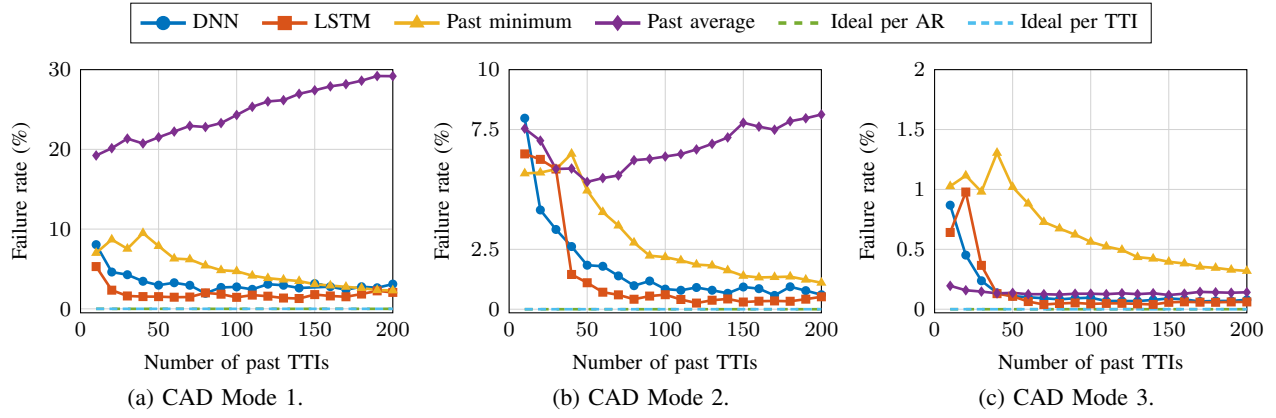

\begin{table}[t]\footnotesize
\centering
\caption{Performance Summary of DNN and LSTM Channel Predictors against benchmarks.}
\label{tab:res_ML_summary_perc}
\renewcommand{\arraystretch}{1.3} 

\begin{tabular}{|p{1.5cm}|>{\centering\arraybackslash}p{0.7cm}|>{\centering\arraybackslash}p{0.7cm}|>{\centering\arraybackslash}p{0.7cm}|>{\centering\arraybackslash}p{0.7cm}|>{\centering\arraybackslash}p{0.7cm}|>{\centering\arraybackslash}p{0.7cm}|}
\hline

& \multicolumn{3}{c|}{$\mathbf{S}(\%)$} & \multicolumn{3}{c|}{$\mathbf{R_{sf}(\%)}$}  \\
\cline{2-7} 

\textbf{Method} & \textbf{10 \acp{TTI}} & \textbf{50 \acp{TTI}} & \textbf{100 \acp{TTI}} & \textbf{10 \acp{TTI}} & \textbf{50 \acp{TTI}} & \textbf{100 \acp{TTI}} \\
\hline \hline

DNN (MAE) & 60.32 & 58.14 & 58.09 & 0.89 & 0.31 & 0.26  \\\hline
DNN (MSE) & 58.69 & 56.85 & 56.99 & 0.67 & 0.36 & 0.33  \\\hline
DNN (MAE+MSE) & 54.39 & 52.24 & 51.52 & 0.73 & 0.24 & 0.15  \\\hline
LSTM (MAE) & 59.01 & 54.27 & 54.49 & 0.80 & 0.25 & 0.21  \\\hline
LSTM (MSE) & 58.13 & 53.9 & 54.39 & 0.68 & 0.22 & 0.19  \\\hline
LSTM (MAE+MSE) & 53.08 & 50.66 & 48.69 & 0.75 & 0.13 & 0.13  \\\hline 
Past minimum & 60.11 & 50.55 & 46.35 & 0.94 & 0.89 & 0.47  \\\hline
Past average & 73.03 & 73.11 & 72.98 & 2.42 & 2.85 & 3.44  \\\hline
Ideal per AR & \multicolumn{3}{c|}{60.43} & \multicolumn{3}{c|}{0} \\\hline
Ideal per TTI & \multicolumn{3}{c|}{75.69} & \multicolumn{3}{c|}{0} \\\hline

\end{tabular}
\end{table}

\subsection{Channel Estimation using LSTM}

\removedAmorosa{The second group of results refer to the LSTM specified in Section~\ref{ml-pred}, with the same loss functions used for the DNN.}

\subsubsection{LSTM with MAE}

\changedAmorosa{
Fig. \ref{fig:DNN-metrics-MAE} report results for the \ac{LSTM} (along with \ac{DNN}) when considering  \ac{MAE} as loss function. 
Fig. \ref{fig:DNN-metrics-MAE}, in particular, refers to the aggregated CAD Modes. The observations that can be made are comparable to those stated in the previous cases.
We observe that the \ac{LSTM} allows to achieve a score close to the \textit{Ideal per AR} and such trend is not sensitive to the $N$ choice (see Fig.~\ref{fig:DNN-path-MAE}). However, we also note that the score is lower than the one achieved with the \ac{DNN}. 
Also in this case, for values of $N$ greater than 50 past \acp{TTI}, the SLA failure is below 0.005 (see Fig.~\ref{fig:DNN-tot-fail-MAE}). 
}{As illustrated in the aggregated metrics of Fig.~\ref{fig:DNN-fail-cad-MAE}, the recurrent model tracks the \textit{Ideal per AR} score closely across all values of $N$, though maintaining a slightly lower nominal profile than the \ac{DNN}. Reliability convergence matches the feed-forward model, securing an SLA failure rate below $0.5\%$ when $N \geq 50$.}

\changedAmorosa{Fig. \ref{fig:DNN-fail-cad-MAE} shows the SLA failure rates per each individual \ac{CAD} Mode. 
Also in this case,  observations similar to the \ac{DNN} case hold. In general the \ac{LSTM} proves to perform better than the benchmarks with a window of past \acp{TTI} greater than 50. Additionally, we can observe that the failure rates achieved by the \ac{LSTM} tend to be lower compared to the \ac{DNN} counterpart. In particular, considering the minimum SLA failure rate for each \ac{CAD} Mode, we note: for \ac{CAD} Mode 1 a minimum equal to $1.94\,10^{-2}$ for the \ac{DNN} and to $1.26\,10^{-2}$ for the \ac{LSTM}; for \ac{CAD} Mode 2 a minimum equal to $5.76\,10^{-3}$ for the \ac{DNN} and to $2.56\,10^{-3}$ for the \ac{LSTM}; for \ac{CAD} Mode 3 a minimum equal to $6.47\,10^{-4}$ for the \ac{DNN} and to $3.94\,10^{-4}$ for the \ac{LSTM}. The overall minimum SLA failure rate indipendently from the \ac{CAD} Mode is $2.11\,10^{-3}$ for the \ac{DNN} and to $1.91\,10^{-3}$ for the \ac{LSTM}.}{At the individual mode level (Fig.~\ref{fig:DNN-fail-cad-MAE}), the \ac{LSTM} reveals distinct performance advantages over the \ac{DNN} when $N \geq 50$. The minimum recorded failure rate drops from $1.94\%$ to $1.26\%$ in \ac{CAD} Mode 1, from $0.58\%$ to $0.26\%$ in \ac{CAD} Mode 2, and from $0.065\%$ to $0.039\%$ in \ac{CAD} Mode 3.} \addedAmorosa{The net global minimum failure rate drops to $0.19\%$ for the LSTM compared to $0.21\%$ for the DNN}. 
\changedAmorosa{However, the slight improvement obtained with the \ac{LSTM} comes at the cost of a longer training time compared to the \ac{DNN}, which performs comparably in terms of SLA failure rates.   ({Qui purtroppo non abbiamo una misura vera e propria della differenza di tempo. Da valutare se tenere questa frase)}}{This incremental gain in safety margins stems from the sequential processing capability of the \ac{LSTM}, though the architectural choice between the two models represents a direct trade-off between strict safety optimization and computational overhead.}

\subsubsection{LSTM with MSE \addedAmorosa{and Mixed Asymmetric Losses}}

\addedAmorosa{As shown in Table~\ref{tab:res_ML_summary_perc}, the \ac{LSTM}-MSE model displays tracking patterns close to its MAE baseline. However, when executing under the asymmetric MAE+MSE loss, the \ac{LSTM} delivers the most robust metrics of the study: at $100$ \acp{TTI}, it reduces the global SLA failure rate to an absolute low of $0.13\%$, proving highly responsive to overestimation penalties.}

\subsection{Main Findings on the Short-Term Channel Prediction}\label{find}

The conducted study provides several insights about the behavior of the covered solutions. For what concerns the short-term channel prediction based only on the history, the following main considerations hold:
\begin{itemize}
    \item \textit{Past average} provides the highest score among the non-ideal predictions, but the highest SLA failure rate. 
   The larger the number of the past TTIs ($N$), the larger the 
    SLA failure rate, the lower the path score. 
    \item \textit{Past minimum} reduces the SLA failure rate 
    as the number of past TTIs ($N$) increases, but significantly reduces the path score, while a larger $N$ improves it. 
\end{itemize}

For \ac{ML}-based short-term channel prediction, the main findings are as follows:
\begin{itemize}
\item With \ac{DNN}, the score only slightly decreases as the number of considered \acp{TTI} $N$ increases.
\item After 50 past \acp{TTI}, the SLA failure rate converges, and the best performance is achieved. 
\item An SLA failure rate equal to 0 is not achieved.
\item No remarkable differences are observed with different models (\ac{DNN} or \ac{LSTM}) or loss functions (MAE, MSE, and combination of MAE and MSE).
\end{itemize} 

The results are overall very promising and require further validation. The DNN appears sufficient with the current modeling, and future work should be dedicated to improve such modeling and to investigate if the approach applies when changing the scenario and environmental conditions.

\section{Conclusion and Future Work}\label{sec:concl}

\addedAmorosa{In this paper, we introduced an ML-driven methodology for proactive URLLC service adaptation in 5G/6G connected vehicular networks. By utilizing accurate channel datasets generated via Sionna-RT and SUMO in a  realistic urban topology, we tested the ability of DNN and LSTM architectures to forecast short-term channel degradation. The results confirm that ML-based prediction models significantly outperform standard history-based conservative short-term channel estimation. Specifically, considered ML models utilizing a custom asymmetric MAE/MSE loss function effectively restricts SLA failure rates to strictly bounded URLLC margins while maintaining an operational path score close to the ideal, perfect-knowledge upper bound.}
Future work will target scenarios with multiple vehicles and interference to assess to which extent the prediction can be affected. 

\section*{Acknowledgment} 
This work has been conducted in the framework of the CNIT-WiLab and the WiLab-Huawei Joint Innovation Center.


\bibliographystyle{IEEEtran}
\bibliography{IEEEbstctl,biblio_papers,biblio_standard}

\end{document}